\documentstyle[12pt]{article}
\newcommand{\lsim}{\mbox{ $
{}_{{}_{\textstyle\sim}}  \! \! \! \! \! {}^{{}_{\textstyle<}}$}}
\begin{document}
\renewcommand{\thefootnote}{\fnsymbol{footnote}}
\thispagestyle{empty}
\vspace*{-1 cm}
\hspace*{\fill}  \mbox{WUE-ITP-95-004} \\
\hspace*{\fill}  \mbox{UH-511-823-95} \\
\vspace*{1 cm}
\begin{center}
{\Large \bf Production of Supersymmetric Higgs Bosons \\ at LEP $\otimes$ LHC
\\ [3 ex] }
{\large F.~Franke\footnote{ email: fabian@physik.uni-wuerzburg.de}
\\ [2 ex]
Institut f\"ur Theoretische Physik, Universit\"at W\"urzburg \\
D-97074 W\"urzburg, Germany
\\ [3 ex]
T. W\"ohrmann\footnote{
email: thomas@uhhepg.phys.hawaii.edu}\\ [2 ex]
Department of Physics and Astronomy, University of Hawaii \\
Honululu, Hawaii 96822, USA}
\end{center}
\vfill

{\bf Abstract}

Within the Minimal Supersymmetric Standard Model (MSSM), we study
the production of the neutral scalar and pseudoscalar as well as
the charged Higgs bosons together with fermions or sfermions
in deep inelastic $ep$ scattering at $\sqrt{s}=1.6$ TeV.
We focus on the parameter space where a Higgs particle is
likely to be invisible at LEP2 and LHC. Although we choose
gaugino/higgsino mixing scenarios that maximize the corresponding
production rates we find only for the production of the scalar Higgs bosons
in the non-supersymmetric channels
non-negligible cross sections of the order of $10^2$ fb.
\vfill
\begin{center}
March 1995
\end{center}

\section{Introduction}
After the observation of the top quark \cite{top}
the Higgs boson
remains the last experimentally undetected component of the
standard model (SM).
The discovery of a Higgs particle could
not only lead to a deeper understanding of the origin of mass and the
nature of electroweak symmetry breaking in the framework of
the SM, but
may also provide an essential hint for models with an enlarged Higgs sector.
Supersymmetric theories offer the most attractive possibility for
such extended models.
While in the SM there exists only one scalar Higgs boson, the
Minimal Supersymmetric Standard Model (MSSM) contains two
Higgs doublets with vacuum expectation values $v_1$ and $v_2$
($\tan\beta=v_2/v_1$) leading to five physical Higgs bosons,
namely two neutral scalars $h$ and $H$ ($m_H > m_h$), one
neutral pseudoscalar $A$ and a pair of degenerate charged Higgs bosons $H^\pm$.
The masses of the scalar Higgs particles and their mixing angle $\alpha$ can
be parametrized in terms of two parameters which are normally chosen
to be $m_A$ and $\tan\beta$.

With the next generation of high energy colliders the search for the
Higgs boson in the SM as well as in the MSSM enters a new
phase. Unfortunately, for both models the proposed new
colliders LEP2 and LHC may probably leave a significant
region of the possible Higgs masses uncovered \cite{janot}.
Therefore one has to think of further colliders in order to be
sensitive to the full parameter space of the Higgs sector.
The  next possible step could be the construction of a new
$ep$ machine by combinating LEP2 and LHC.
In the SM, the cross section for Higgs production by
electron proton scattering is known to be of the
order of $10^2$ fb \cite{zeppenfeld,bluemlein},
therefore being on the edge of the
discovery limit.
In this letter we want to
study the prospects for discovering a MSSM Higgs boson at
LEP $\otimes$ LHC with a center of mass energy $\sqrt{s}=1.6$ TeV.

The present experimental mass limits for a Higgs boson are based on the
negative Higgs search at LEP1. For the SM Higgs boson the mass is
found to be larger than 63.5 GeV \cite{alephsm} while the lower mass
bounds for the scalar and pseudoscalar Higgs particles in the MSSM are
44 GeV ($\tan\beta > 1$) and 21 GeV, respectively \cite{alephmssm}.
The resulting excluded parameter space in the $(m_A,\tan\beta)$ plane
of the MSSM is given in \cite{janot}
together with the accessible domains by LEP2 and LHC.
Here, identification of a Higgs boson is expected to be possible by
the decay channels $h,H \longrightarrow ZZ \longrightarrow 4$ leptons
for $m_{h,H} > 2m_Z$,
or $h,H \longrightarrow \gamma \gamma$.
While the 4-lepton channel provides a clear signal easy to classify,
the 2-photon final state also leads to a favorable
signature but is mostly heavily suppressed. We focus in our
analysis of Higgs production at $ep$ colliders on the
region $100$ GeV $<m_A<200$ GeV, $5<\tan\beta<10$,
where the Higgs boson dominantly decays into $b\bar{b}$
pairs that are difficult to discriminate from QCD background.
For larger values of $\tan\beta$ the Higgs mass range uncovered
by LEP2 and LHC significantly shrinks.

Production of a SM Higgs in $ep$ scattering in the corresponding mass
region $80$ GeV $\lsim m_h \lsim 180$ GeV
by $W$ fusion including QCD and QED corrections was
recently studied by Bl\"umlein et al. \cite{bluemlein}.
We extend this analysis to scalar, pseudoscalar and charged Higgs bosons
of the MSSM. Since in \cite{bluemlein} the first order QCD and
leading QED corrections were found to be as small as
of the order of a few percent
of the total cross section, we restrict ourselves to the
Born level, but consider all Higgs-fermions as well as
Higgs-sfermions production channels.
As it turns out, in the parameter space inaccessible to LEP2/LHC,
only the cross sections for the production of the neutral scalars reach
sizeable values. But they maximally reach the SM values so that the
analysis of Ref. \cite{bluemlein} is applicable, or they are too small
for a reliable identification of the Higgs.
In every case, an $ep$ collider LEP $\otimes$ LHC does not
seem well suited for the Higgs search in the MSSM in the parameter space
not covered by LEP2 and LHC. So an $e^+e^-$ collider with
a center-of-mass energy of 500 GeV appears to be absolutly
necessary in order to explore the full MSSM parameter space
in the Higgs sector.

This paper is organized as follows: In Sec.~2 we present the different
production channels and
choose typical scenarios. The discussion of the numerical results
follows in Sec.~3

\section{Production channels and scenarios}
Higgs boson production by deep inelastic $ep$-scattering takes
place in processes with either a lepton and a quark or a scalar lepton and
a scalar quark
in the final state. The corresponding
Feynman graphs are shown in Fig.~1.

The neutral scalar Higgs bosons are produced together with both
fermions or sfermions where all Feynman graphs of Fig.~1
contribute. Here, associated Higgs-neutrino production proceeds
via $W$-fusion, while $Z$-fusion leads to associated Higgs-electron
production (Fig.~1 (a)).
The cross sections are
related to those in the standard model \cite{zeppenfeld,bluemlein} by
\begin{eqnarray}
\sigma (ep\rightarrow h lqX) & = & \sin^2 (\alpha-\beta)
\sigma (ep\rightarrow H_{SM} lqX)  \\
\sigma (ep\rightarrow H lqX) & = & \cos^2 (\alpha-\beta)
\sigma (ep\rightarrow H_{SM} lqX)
\end{eqnarray}
where $\alpha$ is the mixing angle between the two Higgs states.

Production of a neutral Higgs boson with two sfermions
is described by the Feynman graphs in Fig.~1 (b) - (d).

The neutral pseudoscalar Higgs and the charged Higgs bosons can
be produced only in the supersymmetric channels
$ep \rightarrow A \tilde l \tilde q X,$ $
H^{\pm} \tilde e \tilde q X,$ $ H^- \tilde \nu \tilde q X$.
However, not all supersymmetric graphs in Fig.~1 (b) - (d) contribute to
all of these processes.
So the production of the pseudoscalar Higgs is only mediated via
neutralino/chargino fusion (c) when the Yukawa couplings of the first two
generations are neglected.

The scenarios
for the numerical calculation comply with two requirements.
First, Higgs bosons are not detectable neither at LEP2 nor at
LHC. Therefore we vary the mass of the pseudoscalar Higgs from
100 to 200 GeV at $\tan\beta=5$.
These parameters cover all for the Higgs production
relevant values of the Higgs masses (Fig.~2) and of the reduction factor
$\sin^2(\alpha-\beta )$ (Fig.~3). Radiative corrections to
the Higgs masses and mixings due to top/stops loops are included
with $A_t=0$ GeV and the stop masses
$m_{\tilde t_1}=150$ GeV, $m_{\tilde t_2}=500$ GeV.

Second, the parameters of the gaugino/higgsino sector
lead to maximal production rates in the channels with sfermions.
Generally this happens if the lightest neutralino
assumed to be the lightest supersymmetric particle (LSP)
is a mixture of
higgsino and zino eigenstate. So we choose for the $SU(2)$ gaugino
mass
$M=220$ GeV and for the parameter in the superpotential
$\mu =160$ GeV leading to a LSP with a mass of 84 GeV.
As usual we
assume for the $U(1)$ gaugino mass in the neutralino mixing
matrix the GUT relation $M'=\frac{5}{3} \tan^2 \theta_W M\approx 0.5 M$.

All calculations are performed
for a center of mass energy $\sqrt{s}=1.6$ TeV and slepton and squark
(apart from the stop) masses
$m_{\tilde{l}}=100$ GeV, $m_{\tilde{q}}=200$ GeV well beyond the current
experimental limits.

For the production of massless fermions ($ep\rightarrow hlqX,HlqX$)
the squared momentum transfer at the quark vertex $\tilde Q^2$
entering the quark distribution functions $q(\tilde Q^2,x)$ is given by
$Q^{2}_{hadr.}=-(p_{q_{out}} -p_{q_{in}})^2$. In this case we
impose a cut for $\tilde Q^2$ with
$Q^{2}_{cut}=0.25 (\mbox{GeV})^2$, which is the lower bound
for $\tilde Q^2$ leaving the quark distribution functions \cite{GRV}
still valid. For the
production of massive sfermions we do not need such a cut. But on the other
side the calculation of the squared momentum transfer
$\tilde Q^2$ in the quark distribution
functions becomes more complicated. For the
Feynman graphs (b) and (d) this momentum transfer is given by
$\tilde Q^2 = Q^{2}_{hadr.}=
-(p_{\tilde q} -p_q)^2$, whereas in case of graph (c) $\tilde Q^2$ is given by
$Q^{2}_{lept.}=-(p_{\tilde l} -p_e)^2$. All calculations are based on the
formula \cite{boro}
\begin{eqnarray}
\sigma &=& \int |{\cal M}_b +{\cal M}_d|^2 q(Q^{2}_{hadr.},x)+
2\mbox{Re}({\cal M}_b {\cal M}_{c}^{\ast} + {\cal M}_d {\cal M}_{c}^{\ast})
\sqrt{q(Q^{2}_{hadr.},x)} \sqrt{q(Q^{2}_{lept.},x)}
\nonumber \\ & & +|{\cal M}_c |^2
q(Q^{2}_{lept.},x) \, \mbox{d} Lips \, \mbox{d} x,
\end{eqnarray}
For the remaining parameters of the SM we use $\sin^2 \theta_W =0.228 $
and $m_Z =91.2$ GeV.
\section{Numerical results}
The results for the production of the neutral scalar Higgs bosons
together with neutrinos or electrons are shown in Fig.~4. Since
both
the masses of the Higgs bosons
and the parameter $\sin^2 (\alpha -\beta )$ influence
significantly the cross sections, the results have to be
interpreted with the help of Figs.~2 and 3.
In the case of the production
of the lighter Higgs boson $h$ the cross section increases between
$m_A=100$ GeV and 200 GeV due to the increasing values of
 $\sin^2 (\alpha -\beta )$
while the mass of the lighter Higgs boson is
nearly constant in this range ($m_{h}=85$ GeV for $m_A=100$ GeV
and $m_{h}=98 $ GeV for $m_A=200$ GeV). For large values of the
pseudoscalar Higgs mass it approaches the respective results of the
SM.
The strong decrease of the cross section for $H$ as a function of
$m_A$ is a result of both
the increase of its mass and the decrease of
$\cos^2 (\alpha -\beta )$ approaching 0
for large values of $m_A$.

Generally, the cross sections for Higgs-neutrino production
dominate over those for Higgs-electron production, but
both are at the order of $10^2$ fb and close to their
standard model values \cite{bluemlein}.

In Table 1 we give some typical results for Higgs-sfermion production. Even in
this most optimistic supersymmetric scenario
the cross sections are all smaller than $2 \cdot 10^{-4}$
pb. The main contributions are given by the Feynman graph
Fig.~1~(d) with neutralino/chargino fusion
whereas the graphs Fig.~1~(b) and (c) are of minor importance.
Since our scenario with the LSP being a higgsino/gaugino mixture
maximizes the total cross section and
especially also the contribution from graph (d),
it suppresses the contributions from graphs (b) and (c).
The largest cross sections of order $10^{-1}$ fb
are obtained for the production of a sneutrino
together with any kind of Higgs or for
the associated production of charged Higgs
bosons arising by strong chargino couplings. Especially the $\tilde \nu
H^- $-production is that one with the biggest cross section due
to the strong chargino couplings at the leptonic vertex. On the other side
we get the smallest cross sections for the associated production of
selectrons and heavy neutral Higgs bosons (scalar as well as pseudoscalar).
Even for our rather small values of $m_{\tilde l}$ and $m_{\tilde q}$,
the large total sum of the masses of the particles produced
in the supersymmetric channels ($\geq 400$ GeV)
generally leads to these small cross sections.

\section{Conclusion}
Cross sections for the production of the neutral scalar and
pseudoscalar as well as the charged Higgs bosons by
deep inelastic $ep$ scattering with $\sqrt{s}=1.6$ TeV were
computed within the MSSM. We focussed on that part of the
MSSM parameter space inaccessible both to LEP2 and
to LHC. While the pseudoscalar and charged Higgs bosons can be
produced only together with sfermions with
negligibly small cross sections, the associated
scalar Higgs-neutrino production reaches sizeable cross sections
of the order of $10^2$ fb comparable to the standard model and leads to the
typical signature $b\bar b \not \! E$.
The prospects for detecting a Higgs boson at LEP $\otimes$ LHC,
however, strongly depend on the hadronic background.
This background arises mainly by the basic subprocess $eg\rightarrow eb\bar b$
where the collinear electron escapes detection. Cuts on transverse missing
Energy $ \not \! E_T $ and detailed Monte Carlo studies are necessary to
answer the question whether this background is reducible. Another interesting
signature of Higgs production could arise by the Higgs decay into a $\tau $
pair, with a rather small background from the subprocess
$ eq\rightarrow \nu Z X,\, Z\rightarrow \tau \bar \tau$ \cite{zep2}. For
this signature, however, a larger luminosity
than $10^3 \mbox{pb}^{-1}/\mbox{year}$ as expected
for LEP $\otimes $ LHC may be necessary.

\section*{Acknowledgements}
We would like to thank
X.~Tata and H.~Fraas for many helpful discussions and their careful
readings of the manuscript. F.~F.~gratefully acknowledges support
by Cusanuswerk,
T.~W.~is supported by Deutsche Forschungsgemeinschaft and, in part,
by the U.S. Department of Energy Grant No. DE-FG-03-94ER40833.

\newpage
\section*{Figure Captions}
\begin{enumerate}
\item Feynman graphs for the processes $ep\rightarrow hlqX,\, HlqX$ (a) and
$ep\rightarrow h\tilde l \tilde q X,$ $ H\tilde l \tilde q X,$ $
A\tilde l\tilde q X,$ $H^{\pm} \tilde e \tilde q X $, $ H^-
\tilde \nu \tilde q X$ (b), (c), (d).
\item Masses of the two neutral scalar Higgs bosons $h$, $H$
as a function of $m_A$ for $\tan\beta=5$, $A_t=0$ GeV,
$m_{\tilde{t}_1}=150$ GeV, $m_{\tilde{t}_2}=500$ GeV.
\item $\sin^2 (\alpha -\beta )$ as a function of $m_A$
for $\tan\beta=5$, $A_t=0$ GeV,
$m_{\tilde{t}_1}=150$ GeV, $m_{\tilde{t}_2}=500$ GeV.
\item Cross sections for the associated Higgs-lepton production
with $\tan\beta=5$, $A_t=0$ GeV,
$m_{\tilde{t}_1}=150$ GeV, $m_{\tilde{t}_2}=500$ GeV.
\end{enumerate}

\vspace{2cm}

\section*{Table Caption}
\begin{enumerate}
\item Cross sections for $m_A=100$ GeV and $m_A=200$ GeV
for the associated Higgs-slepton production with
$M=220$ GeV, $\mu = 160$ GeV,
$\tan\beta=5$, $A_t=0$ GeV.
The results are obtained with slepton and squark (apart from the stop)
masses $m_{\tilde{l}}=100$ GeV, $m_{\tilde{q}}=200$ GeV and
stop masses $m_{\tilde{t}_1}=150$ GeV,
$m_{\tilde{t}_2}=500$ GeV.
\end{enumerate}

\newpage
\vspace*{5cm}
\begin{center}
\begin{tabular}{|c||c|c|}
\hline & $m_A = 100 $ GeV & $ m_A = 200 $ GeV \\ \hline
$ ep\rightarrow h \tilde \nu \tilde q X $ & 0.03 fb & 0.06 fb \\ \hline
$ ep\rightarrow H \tilde \nu \tilde q X $ & 0.1 fb & 0.03 fb \\ \hline
$ ep\rightarrow h \tilde e \tilde q X $ & 0.01 fb & 0.01 fb \\ \hline
$ ep\rightarrow H \tilde e \tilde q X $ & 0.007 fb & 0.002 fb \\ \hline
$ ep\rightarrow H^- \tilde \nu \tilde q X $ & 0.18 fb & 0.08 fb \\ \hline
$ ep\rightarrow H^- \tilde e \tilde q X $ & 0.08 fb & 0.03 fb \\ \hline
$ ep\rightarrow H^+ \tilde e \tilde q X $ & 0.035 fb & 0.01 fb \\ \hline
$ ep\rightarrow A \tilde \nu \tilde q X $ & 0.055 fb & 0.02 fb \\ \hline
$ ep\rightarrow A \tilde e \tilde q X $ & 0.006 fb & 0.002 fb \\ \hline
\end{tabular}
\\
\vspace{0.5cm}
{\Large Table 1 }
\end{center}

\begin{thebibliography}{99}
\bibitem{top} CDF Collaboration, F. Abe et al., FERMILAB-PUB-95/022-E,
hep-ex/9503002; \\
D0 Collaboration, S. Abachi et al., FERMILAB-PUB-95/028-E, hep-ex/9503003
\bibitem{janot} P. Janot,
Proceedings of the Workshop on Physics and Experiments with
Linear $e^+e^-$ colliders, Vol.~I, p.~192, World Scientific,
Eds. F.A. Harris, S.L. Olsen, S. Pakvasa, X. Tata
\bibitem{zeppenfeld} G. Grindhammer, D. Haidt, J. Ohnemus,
J. Vermaseren and D. Zeppenfeld,
Proceedings of the EFCA Large Hadron Collider Workshop, Aachen, 1990,
CERN90-10, ECFA 90-133, Vol.~II, p.967, Eds. G. Jarlskog and D. Rein,
\bibitem{bluemlein} J. Bl\"umlein, G.J. van Oldenborgh and R. R\"uckl,
Nucl. Phys. {\bf B 395} (1993) 35
\bibitem{alephsm} G.~Gopal, presented at the Aspen Winter Conference on
Particle Physics Beyond the Year 2000, Aspen, CO, January 1994.
\bibitem{alephmssm} ALEPH Collaboration, D. Buskulic et al.,
Phys. Lett. {\bf B 313} (1993) 312
\bibitem{GRV} M.~Gl\"uck, E.~Reya, A.~Vogt, Z.~Phys.~C {\bf 53} (1992) 127
\bibitem{boro} M.~B\"ohm, A.~Rosado, Z.~Phys.~C {\bf 34} (1987) 117
\bibitem{zep2} U. Baur, B.A. Kniehl, J.A.M Vermaseren and D. Zeppenfeld,
Proceedings of the EFCA Large Hadron Collider Workshop, Aachen, 1990,
CERN90-10, ECFA 90-133, Vol.~II, p.956, Eds. G. Jarlskog and D. Rein
\end{thebibliography}
\end{document}